# Two-dimensional layered materials meet perovskite oxides: A combination for high-performance electronic devices

Allen Jian Yang[1,*], Su-Xi Wang[2], Jianwei Xu[2,3,4], Xian Jun Loh[2,3,4], Qiang Zhu[2,3,5], Xiao Renshaw Wang[1,6,*]

[1]Division of Physics and Applied Physics, School of Physical and Mathematical Sciences, Nanyang Technological University, 21 Nanyang Link, Singapore 637371
[2]Institute of Materials Research and Engineering (IMRE), Agency for Science, Technology and Research (A*STAR), 2 Fusionopolis Way, Innovis, #08-03, Singapore 13863
[3]Institute of Sustainability for Chemicals, Energy and Environment (ISCE[2]), Agency for Science, Technology and Research (A*STAR), 1 Pesek Road, Jurong Island, Singapore, 627833
[4]Institute of Materials Research and Engineering (IMRE), Agency for Science, Technology and Research (A*STAR), 2 Fusionopolis Way, Innovis, #08-03, Singapore 13863
[5]School of Chemistry, Chemical Engineering and Biotechnology, Nanyang Technological University, 21 Nanyang Link, Singapore 637371
[6]School of Electrical and Electronic Engineering, Nanyang Technological University, 50 Nanyang Ave, 639798, Singapore

*e-mail: yangjian@ntu.edu.sg; renshaw@ntu.edu.sg

**Abstract**

As the Si-based transistors scale down to atomic dimensions, the basic principle of current electronics, which heavily relies on the tunable charge degree of freedom, faces increasing challenges to meet the future requirements of speed, switching energy, heat dissipation, packing density as well as functionalities. Heterogeneous integration, where dissimilar layers of materials and functionalities are unrestrictedly stacked at an atomic scale, is appealing to next-generation electronics, such as multi-functional, neuromorphic, spintronic and ultra-low power devices, because it unlocks technologically useful interfaces of distinct functionalities. Recently, the combination of functional perovskite oxides and the two-dimensional layered materials (2DLMs) led to unexpected functionalities and enhanced device performance. In this review, we review the recent progress of the heterogeneous integration of perovskite oxides and 2DLMs from the perspectives of fabrication and interfacial properties, electronic applications, challenges as well as outlooks. In particular, we focus on three types of attractive applications, namely field-effect transistors, memory, and neuromorphic electronics. The van der Waals integration approach is extendible to other oxides and 2DLMs, leading to almost unlimited combinations of oxides and 2DLMs and contributing to future high-performance electronic and spintronic devices.

Processing, storing, and transmitting information with high efficiency form the basis of the current information age. These tasks, from a fundamental point of view, are performed by modulating the electrical, magnetic, and optical properties of materials and structures with respective excitations. For over five decades, silicon metal-oxide-semiconductor field-effect transistors (MOSFETs) and integrated circuits (ICs) have dominated digital chips, including microprocessors and memories, because they offer continuously increasing performance and decreasing costs. Such a trend has been driven by the celebrated Moore's law, which prescribes an exponential increase of the transistor number, hence the computing power, on a single chip.[1] The transistors for digital computation are required to be fast, low-power switches that can be scaled up in integration density. This has been made possible by continuously reducing the supply voltages and miniaturizing the size of silicon transistors through various material and structural innovations. For example, the adoption of high-$\kappa$ dielectrics and metal gate to replace the original $SiO_2$ and poly-Si gate, respectively, enabled the expansion to the 45-nm technology node.[2] The structural innovation is represented by the transformation from planar to 3D geometries, *e.g.* FinFETs, when the planar miniaturization was severely hindered by the short-channel effects.[3] For a similar reason, the semiconductor technology is now turning to gate-all-around (GAA) transistors, where multiple nanosheet channels are stacked vertically, for

sub-5-nm technology nodes.[4] As for the data storage, there have been several revolutions in the principles as well as in the materials. Some of the most widely adopted technologies are optical discs based on reading and changing the reflectivity of encoding medium, hard disk drives based on sensing magnetic fields through various magnetoresistive effects, and flash memory based on charge-trapping transistors.[5] These innovations in the design and fabrication of devices and circuits, along with the advancements in data communication technologies, enable portable computers and mobile phones connected by the internet, which shape the current everyday life.

With the advent of Big Data, Internet of Things (IoT), artificial intelligence (AI), wearable electronics, and other emerging applications, there comes an enormous demand for the computing power in different forms. However, the further downscaling of silicon transistors is becoming increasingly difficult and costly, and will inevitably reach its fundamental limits as the channel length approaches atomic scales.[6, 7] As well as that, the heating problem has been limiting the clock rates of microprocessors to several gigahertz since 2004.[6] In addition, with the advancement of processor speed and memory size, the delay due to the movement of data between the processor and the memory is becoming the bottleneck of overall computer performance, termed as von Neumann bottleneck,[8] especially when handling large amounts of data. To tackle these challenges, a variety of strategies, classified into so-called More Moore, More than Moore, and beyond CMOS, have been proposed and explored.[9] Briefly, More Moore refers to the further extreme miniaturization of CMOS transistors and increasing areal component densities through continued efforts on material and processing innovations.[7] More than Moore involves adding various non-digital functionalities, including radio-frequency electronics, power electronics, sensors and actuators, to the digital processing units, thereby improving the overall performance of the system.[6] Such combinations are frequently achieved by advanced packaging or on-chip heterogeneous integration. Beyond CMOS is the most radical strategy that aims to replace CMOS transistors with other types of switches that are faster and/or more power-efficient as the fundamental information processing devices. A wide range of options, including spin-FET, tunnel FET, negative-capacitance (NC) FET, nano-electro-mechanical (NEM) switches, and memristive devices, are being intensively researched.[10, 11] Moreover, based on them other computing paradigms, such as in-memory computing, may be developed to circumvent the limitations posed by the conventional von Neumann architecture.

Two-dimensional layered materials (2DLMs), represented by graphene, transition metal dichalcogenides (TMDCs), black phosphorus (bP), *etc.*, can be used to push the limits of conventional semiconductor technologies and to pursue unusual functionalities owing to their appealing structural, electronic, and optoelectronic properties.[12] 2DLMs are characterized by the weak van der Waals (vdW) bonding between adjacent layers that allows realization of atomic layers with dangling-bond-free surfaces. The atomic thickness endows 2DLMs with superior electrostatic control, and the dangling-bond-free surface enables them to retain high mobilities even when thinned to a single atomic layer. Thus, the semiconducting TMDCs have been regarded as a promising channel material for the ultrascaled transistors to extend Moore's law.[13-17] Additionally, the low-temperature processability of 2DLMs renders them suitable for monolithic integration with Si electronics, thus combining the outstanding electronic and photonic properties of 2DLMs with the mature Si technology for More than Moore applications.[12, 18] What's more, 2DLMs can be assembled among themselves or with other materials to form a wide range of van der Waals heterostructures with designed or emergent properties that can be exploited for future computing devices.[19-22]

Recently, perovskite oxides, a large family of complex oxides which share crystal structure and processing conditions, have emerged as a versatile enabler to be integrated with 2DLMs. They are a host of a range of exotic electrical and magnetic properties, including high-temperature superconductivity, ferroelectricity, piezoelectricity, colossal magnetoresistance, and many more.[23-25] The incipient ferroelectric, or quantum paraelectric, perovskite oxides, such as $SrTiO_3$, exhibit extremely large dielectric constants that are highly beneficial to field-effect devices. Perovskite ferroelectric materials are renowned for their robust and large remnant polarization, based on which memory devices in different device structures can be designed and fabricated. What's more, the strong correlation-induced high-temperature superconductivity, Mott transition, half metallicity, *etc.* can also be exploited to develop electronic devices. These outstanding properties make it highly desirable to integrate these perovskite oxides with 2DLMs for the following reasons. First, the ultrathin body of 2DLMs make them very sensitive to the adjacent electric and magnetic field created by the perovskite oxides.[26] Second, as we explained above, 2DLMs possess a few important electronic and optoelectronic properties that are absent in perovskite oxides. Third, unlike the pure perovskite oxide heterostructures, where the high-quality interfaces are restricted by lattice mismatch, the 2DLM-perovskite oxide heterostructures, in principle, can be assembled with arbitrary combinations. Kang *et al*. summarized the synergetic effects arising from 2DLM-complex oxide interfaces and focused on the behavior of 2DLMs on different complex oxide substrates.[27] There are also a few reviews concerning the integration of 2DLMs with oxide materials but most of them are limited to a specific functionality, such as ferroelectricity.[28-30] In this Review, we aim to provide a comprehensive survey of the electronic devices based on 2DLM-perovskite oxide heterostructures, ranging from transistors to memory and neuromorphic devices. First, we give a concise overview of the properties of perovskite oxides. Next, we introduce the fabrication techniques for 2DLM-perovskite oxide heterostructures and their interfacial properties. Then, we elaborate the working mechanisms and highlight the representative works on different types of prototype devices grouped into MOSFETs, nonvolatile memory devices, and neuromorphic devices. Last, we outline the challenges of applying these devices to practical products and propose our perspective on future research.

**AN OVERVIEW OF THE PROPERTIES OF PEROVSKITE OXIDES**

Owing to the intricate interplay of lattice, orbital, charge, and spin, perovskite oxides exhibit rich and technologically useful properties, such as large dielectric constants, ferroelectricity, piezoelectricity, ferromagnetism and strongly correlated behaviors including high-temperature superconductivity, Mott metal-insulator transition, colossal magnetoresistance, multiferroics.[23, 31-33] In the ideal perovskite structure illustrated in Figure 1, A-site cations occupy the corners of the cubic unit cell while the B-site cation is at the center and forms a $BO_6$ octahedron with six O anions at the face centers. In reality, however, this structure frequently undergoes a range of distortions, thereby leading to many derivative crystalline phases accompanied by distinct properties. For example, the displacement of the B-site cation in the $BO_6$ octahedra is the origin of the cubic-to-tetragonal phase transition for many ferroelectric and piezoelectric perovskite oxides.[25, 34] Other types of symmetry breaking processes include the rotation or tilting of the $BO_6$ octahedra and their distortions (*i.e.* Jahn-Teller distortion) associated with the electron orbitals.[35] What's more, aside from the basic $ABO_3$ formula, there are a variety of layered perovskite oxides consisting of alternating $ABO_3$ layers separated by other motifs, thus further enriching the crystalline phases and properties.[36]

The perovskite structure is flexible in accommodating a wide range of elements. The A-site cation is usually a rare earth (Sc, Y, or lanthanide) or alkaline earth element (Be, Mg, Ca, Sr, Ba, or Ra) while the B-site cation is usually a transition metal element (*e.g.* Ti, Fe, Co, Mn) with a smaller cationic radius. Thus, the transition-metal-containing perovskite oxides also belong to transition metal oxides. The *d* electrons from the transition metals experience two types of competing forces, namely Coulombic repulsion between electrons that tends to localize individual electrons at atomic lattice sites and hybridization with the oxygen *p* electron states that tends to delocalize the electrons.[24, 37] The subtle balance directly impacts the transitions between different electronic properties, and is believed to underlie exotic metal-insulator transition, colossal magnetoresistance, and high-temperature superconductivity.[38, 39] Spin of localized *d* electrons directly connects to the magnetism of perovskite oxides. Localized spins that reside in different electronic configurations exhibit different kinds of magnetism. When determining the magnetic state of perovskite oxides, Hund's rule is quantitatively applicable. However, significant corrections to the electronic configuration should be considered when applying it. The corrections on the *d* state of electronic configuration are electrostatic splitting due to the electrostatic effect from neighboring transition metal ions and ligand field splitting due to the covalent metal-oxygen bonds.

The perovskite structure is highly tolerant to foreign cations in either A- or B-sites and to structural defects, providing the additional knobs to enhance or even modulate the properties of the perovskite oxides. Furthermore, the interplays of doping effect, defect engineering and breaking of symmetries lead to a large variety of tantalizing properties, such as ferromagnetism in $SrRuO_3$ arising from the breaking of the time-reversal symmetry, ferroelectricity in strained $SrTiO_3$ arising from the breaking of the inversing symmetry, and high-temperature superconductivity in YBa$_2$Cu$_3$O$_X$ arising from the breaking of the gauge symmetry. Figure 1 summarizes the potentially important properties of perovskite oxides that are attractive for electronic devices. As elaborated above, charge, lattice, spin and orbit are four basic factors in forming the functionalities in perovskite oxides. A variety of functionalities are formed based on the four fundamental elements and their interplays. These functionalities can be applied in a wide spectrum of devices. Representative materials related to the functionalities are also presented.

The marriage of perovskite oxides and 2DLMs is highly attractive, owing to the appealing properties of the materials and even potentially unseen correlated quantum functionalities at their interface. Moreover, because perovskite oxides and 2DLMs are two big families of materials with a vast number of compounds and stoichiometries, this integration of the perovskite oxides and 2DLMs offers almost unlimited combinations and eventually possible applications. Indeed, the recent integration of functional perovskite oxides and 2DLMs has shown convincing experimental results useful to both promising next-generation electronic devices competitive or beyond the state-of-the-art ones and manipulatable interfaces with tantalizing opportunities to alleviate the limitations of current technologies.

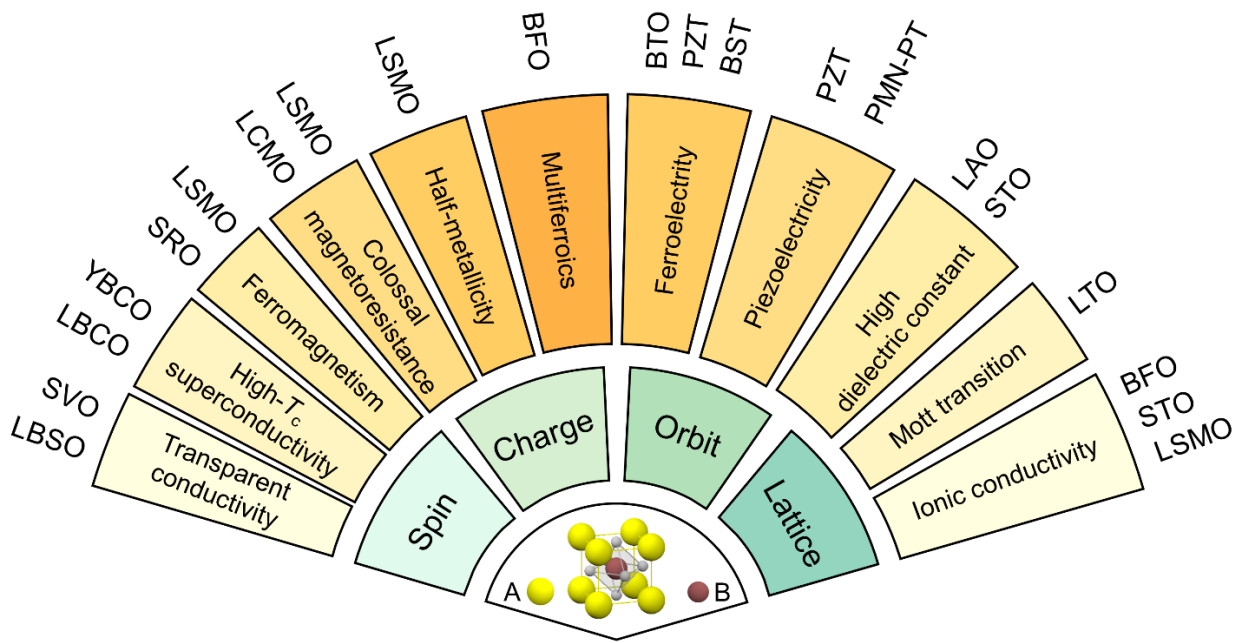

**Figure 1. Technologically important properties of representative perovskite oxides.** LBSO, $La_xBa_{1-x}SnO_3$; SVO, $SrVO_3$; LBCO, $La_{2-x}Ba_xCuO_4$; YBCO, $YBa_2Cu_3O_{7-x}$; SRO, $SrRuO_3$; LSMO, $La_{1-x}Sr_xMnO_3$; LCMO, $La_{1-x}Ca_xMnO_3$; BFO, $BiFeO_3$; BTO, $BaTiO_3$; PZT, $PbZr_xTi_{1-x}O_3$; BST, $Ba_xSr_{1-x}TiO_3$; PMN-PT, $(PbMg_{1/3}Nb_{2/3}O_3)_{1-x}-(PbTiO_3)_x$; LAO, $LaAlO_3$; STO, $SrTiO_3$; LTO, $LaTiO_3$.

## FABRICATION AND INTERFACIAL PROPERTIES OF 2DLM-PEROVSKITE OXIDE HETEROSTRUCTURES

Enabled by the advancements in the growth, transfer, and assembly technologies, a wide range of 2DLM-perovskite oxide heterostructures have been realized. The fabrication process for heterostructures is restricted by the physical and chemical properties of the constituent materials. Since 2DLMs can, in principle, adhere to any surface of interest, the methods of choice for 2DLM-perovskite oxide integration are largely determined by perovskite oxides. Direct growth of crystalline perovskite oxides on 2DLMs is extremely challenging because of the destructive growth conditions for 2DLMs, which involve the high temperature along with the oxidizing atmosphere.[40] For oxide bulk crystals and thin films grown on oxide substrates, the integration is usually a one-way process, which involves transferring 2DLMs onto the oxide surfaces. However, after the realization of freestanding perovskite oxides, the stacking order can be reversed, thus enabling more heterostructure configurations.

Figure 2 illustrates the typical methods for integration between 2DLMs and perovskite oxides. The most straightforward method is mechanical exfoliation of layered crystals, *e.g.* graphite and $MoS_2$ crystals, using an adhesive tape and subsequent transfer of the 2DLMs onto perovskite bulk crystals or thin films grown on lattice-matched perovskite oxide substrates (Figure 2a). In early studies model materials from these two families, *e.g.* graphene and bulk STO or PZT film on Nb-doped STO substrates, were chosen due to their accessibility.[41, 42] Although this method is a versatile and enabling approach to unconventional interfaces, it suffers from low efficiency, location randomness, and small scale. Since then, the dramatic expansion of 2DLM family and the tremendous progress in their transfer technologies[43, 44] have made it possible to prepare 2DLMs on optimal substrates by mechanical exfoliation or chemical synthesis and then transfer them onto perovskite oxide surfaces with an accurate location and in a larger scale (Figure 2b). Consequently, an ever-growing number of 2DLM-perovskite oxide combinations, such as graphene-PMN-PT and bP-STO have emerged.[45, 46] Directly growing 2DLMs on perovskite oxide surface (Figure 2c) has been proven successful for some combinations, such as graphene and $MoS_2$ on STO,[47, 48] but consecutive attempts are scarce probably because the perovskite oxide substrate is not suitable for practical electronics applications.

To integrate perovskite oxides with the technologically relevant platform, tremendous efforts have been devoted to epitaxial growth of perovskite oxides on conventional semiconductor substrates.[49, 50] The functionalities arising from such integration generally have two levels. The first level involves the simple structural integration of the perovskite oxides in the form of a single compound or multiple layers with the desired properties discussed in the previous section. The second level harnesses the complex electrical coupling between perovskite oxides and conventional semiconductors, such as the charge transfer across the heterojunctions and metal-oxide-semiconductor capacitive coupling. These advancements greatly benefit the fabrication of 2DLM-perovskite heterostructures on Si substrates in that 2DLMs can be transferred on this hybrid structure using the similar methods described above (Figure 2d). For example, growth of PZT on Si substrates followed by the transfer of 2DLMs has enabled the fabrication of ferroelectric field-effect transistors (FeFETs) that we will discuss in the following section.[51]

Recently, the realization of freestanding perovskite oxides enables another method for the integration between 2DLMs and perovskite oxides.[52-54] Specifically, by growing the target perovskite oxide films on a lattice-matched oxide sacrificial layer, it is possible to lift off the films and handle them in a manner similar to handling 2DLMs. This method relies on the etching of the sacrificial oxides (Figure 2e), of which selection is restricted by the lattice matching to both the target oxide films and the oxide substrates. The most widely used sacrificial layers include the $(Ca,Sr,Ba)_3Al_2O_6$ family, which is water dissolvable, and certain perovskite oxides, such as LSMO and SRO, that can be selectively etched by cautiously chosen etchants.[53, 55] Benefiting from the freestanding perovskite oxides, the integration between 2DLMs and perovskite becomes more flexible because the stacking orders can be readily controlled and alternating layers of 2DLMs and perovskite oxides are achievable.

Depending on the fabrication methods and constituent materials, the 2DLM-perovskite oxide heterostructures exhibit varied interfacial properties in terms of interfacial bonding and charge transfer. When assembled using the aforementioned transfer techniques, the 2DLMs and perovskite oxides are hold together by the weak van der Waals forces. In these cases, the surfaces of these two types of crystals at the interfaces are relaxed. The two crystals are randomly arranged in relation to each other unless they are deliberately aligned using advanced stacking technologies.[56] Thus, lattice mismatch is not regarded as a limiting factor for the possible material combinations and has negligible effects on the interfacial electronic properties. In contrast, for 2DLMs epitaxially grown on perovskite oxides, the lattice mismatch have a significant influence on the morphology and electronic properties of the resulting 2DLM-perovskite heterostructures.[57] This is exemplified by the chemical vapor deposition (CVD) of graphene and $MoS_2$ on high-$\kappa$ perovskite oxides including STO and LAO.[47, 58-61] The lattice constants of graphene, $MoS_2$, STO, and LAO are 2.468, 3.161, 3.905, and 3.787 Å, respectively. Due to the considerable lattice mismatch, as well as the difference in thermal expansion coefficients, the as-grown 2DLMs are frequently strained. This strain effect may be exploited to engineer or enhance the functionalities of the 2DLMs on perovskite oxides,[62-64] but many further investigations are required to understand and manipulate this effect.

The charge transfer processes in 2DLM-perovskite oxide heterostructures can be generally classified into two types according to the conductivity of the perovskite oxides. In heterostructures formed between considerably conductive perovskite oxides and 2DLMs, which are both abundant in free charge carriers, the charge transfer is largely determined by the Fermi level alignment. Specifically, electrons from the material with a higher Fermi level (smaller work function) transfer to the material with a lower Fermi level (larger work function). This process leads to equal and opposite charge accumulation in both materials, which gives rise to a built-in electric field. A representative example of this type of charge transfer is the one that exists in the $La_{0.7}Sr_{0.3}MnO_3$-$MoS_2$ heterojunction.[65]

La$_{0.7}$Sr$_{0.3}$MnO$_3$ is a metallic perovskite oxide with a relatively large work function (4.8 eV), and MoS$_2$ is an n-type semiconductor with a smaller work function (4.4−4.5 eV) dependent on the substrates[66] from the 2DLM family. In contact, electrons transfer from MoS$_2$ to La$_{0.7}$Sr$_{0.3}$MnO$_3$ and the resulting built-in electric field can be exploited for diode devices.[65] To date, far too little attention has been paid to this type of 2DLM-perovskite oxide heterostructures and their related charge transfer processes, but much more can be expected considering the rich and fascinating functionalities of the conductive perovskite oxides.

Another type of charge transfer exists in the heterostructures formed from 2DLMs and insulating perovskite oxides, which are the main focus of this Review because up to now, the vast majority of reports on 2DLM-perovskite oxide heterostructures deal with insulating perovskite oxides. Similar to the phenomena commonly observed in other 2DLM-insulator systems, the dangling bonds or crystalline defects of these perovskite oxides act as charge traps, resulting in the net charge transfer between 2DLMs and the insulating perovskite oxides at thermal equilibrium.[67] It leads to an effective n-doping (p-doping) effect on the 2DLMs if the trap states are electron donors (acceptors). When the equilibrium is disturbed by the optical illumination or applied gate voltage, the captured charge carriers in the trap states recombine with the excess free carriers with varied dynamics. This charge transfer scenario is frequently interfered by other mechanisms, such as the doping effect of ambient adsorbates and the electric field effect exerted by surface dipole and migrating ions.[68-70] Systematic understanding and reliable control of these interfacial interactions are highly desirable because they play a vital role in the performance of electronic devices that we will discuss in the following sections.

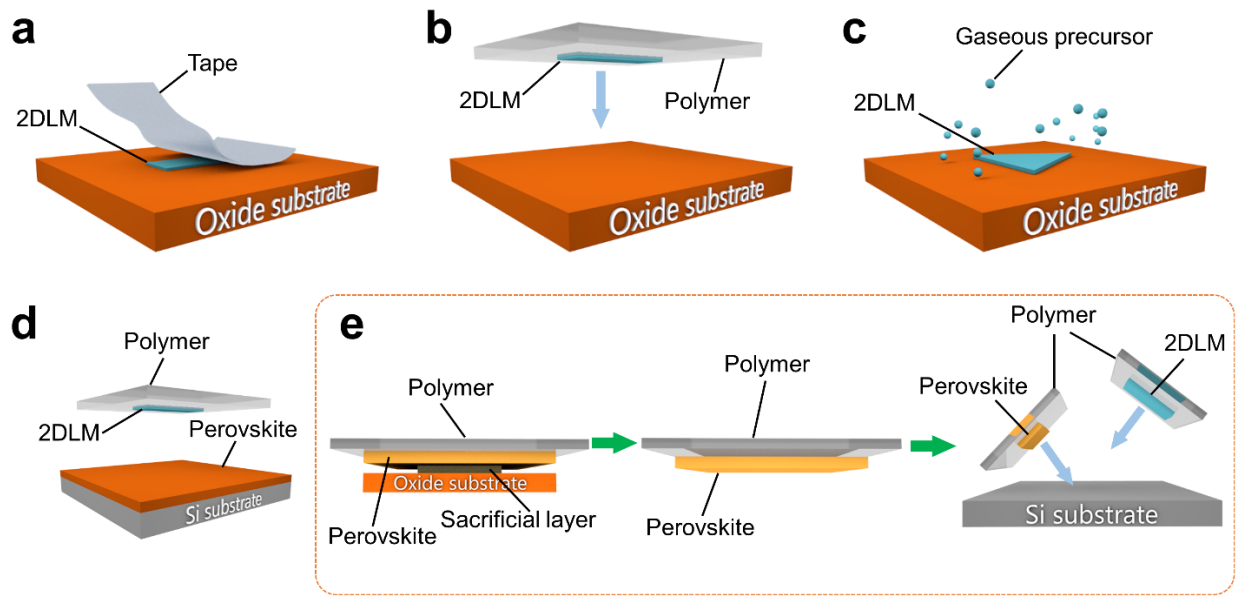

**Figure 2. Integration methods for 2DLM-perovskite oxide heterostructures.** (a) Mechanical exfoliation of 2DLMs on perovskite oxide surface. (b) Transfer of 2DLMs from other substrates onto perovskite oxide surface. (c) Direct growth of 2DLMs on the surface of perovskite oxides. (d) Mechanical exfoliation or transfer of 2DLMs onto perovskite oxides grown on Si substrates. (e) Release of epitaxial perovskite oxides from oxide substrates and successive transfer of 2DLMs and perovskite oxides onto Si substrates to form heterostructures.

**ELECTRONIC DEVICES BASED ON 2DLM-PEROVSKITE OXIDE HETEROSTRUCTURES**

Thanks to richness of the functionalities of the perovskite oxides, a variety of electronic devices have been explored based on the 2DLM-perovskite oxide heterostructures, ranging from MOSFETs to magnetic sensors. Table 1 summarizes the fabrication methods of the 2DLM-perovskite oxide heterostructures, the key performance merits of the devices, and the critical properties the perovskite oxides exhibit in these devices. Albeit with the extensive range of device types, most up-to-date reports are concerned with the MOSFETs, memory devices, and neuromorphic devices, which are the focus of the current Review.

**Table 1.** Summary of the representative electronic devices based on 2DLM-perovskite oxide heterostructures.

| Material | Oxide property | Device | Fabrication | Performance | Reference |
| --- | --- | --- | --- | --- | --- |
| BTO-MoS$_2$ | High-$\kappa$ | n-MOSFET | Method e | Mobility 5−70 cm$^2$ V$^{-1}$s$^{-1}$ | [71] |
| PZT-MoS$_2$ | High-$\kappa$ | n-MOSFET | Method d | EOT ~ 1.1 nm | [72] |
| STO-MoS$_2$ | High-$\kappa$ | n-MOSFET | Method a | $I_{on}/I_{off}$ ~ 10$^7$ | [73] |
| STO-MoS$_2$ | High-$\kappa$ | n-MOSFET | Method e | SS ~ 66 mV dec$^{-1}$ | [74] |
| STO-MoS$_2$ | High-$\kappa$ | n-MOSFET | Method e | CET < 1 nm | [75] |
| STO-WSe$_2$ | High-$\kappa$ | p-MOSFET | Method e | $I_{on}/I_{off}$ > 10$^7$ | [74] |
| BFO-Gr | Tunnel barrier | Gr FET | PLD on Gr | $I_{on}/I_{off}$ ~ 7 × 10$^7$ | [76] |
| PMN-PT-Gr | Ferroelectric | FeFET | Method b | Ferroelectric hysteresis | [45, 77] |
| PZT-BP | Ferroelectric | FeFET | Method d | Polarization-dependent $I_{ph}$ | [78] |
| PZT-Gr | Ferroelectric | FeFET | Method b | Ambipolar conduction | [79] |
| PZT-Gr | Ferroelectric | FeFET | Method d | On/Off window ~ 5.3 | [80] |
| PZT-MoS$_2$ | Ferroelectric | FeFET | Method d | On/Off window ~ 17 | [51] |
| PZT-MoS$_2$/WSe$_2$ | Ferroelectric | FeFET | Method a | On/Off ratio ~ 10$^2$ | [81] |
| BTO-Gr | Ferroelectric | FTJ | Method b | TER ~ 6 × 10$^5$% | [82] |
| BTO-MoS$_2$ | Ferroelectric | FTJ | Method a | TER ~ 10$^4$ | [83] |
| BFO-MoS$_2$ | Ionic conductor, UV absorber | Optoelectronic memory | Method a | Visible write, UV erase | [84] |
| PZT-MoS$_2$ | Ferroelectric | Optoelectronic memory | Method d | Optical write/erase | [51, 85] |
| STO-Gr | UV absorber | Optoelectronic memory | Method a | All-optical manipulation | [86] |
| BTO-MoS$_2$ | Ferroelectric | Artificial synaptic device | Transfer BTO nanoflake | Long-term plasticity | [87] |
| BTO-MoS$_2$ | Ferroelectric | Neuromorphic vision sensor | Method b | Image recognition rate ~ 91% | [88] |
| PZT-WS$_2$ | Ferroelectric | Optoelectronic synapse | Method a | Synaptic plasticity | [89] |
| BTO-Gr | Piezoelectric | Piezoelectric resonator | Method e | Vibration frequency ~ 233 GHz | [90] |
| LSMO-MoS$_2$ | Hole metal | Photodiode | Method b | $V_{OC}$ ~ 0.4 V | [65] |
| BFO-WSe$_2$ | Ferroelectric | P-n homojunction | Method b | Rectifying | [91] |
| LSMO-MoS$_2$ | Ferromagnetic metal | Spin valve | Method c | GMR ~ 0.8% | [92] |
| PZT-MoTe$_2$ | Piezoelectric | Strain effect transistor | Method d | SS ~ 6 mV dec$^{-1}$ | [93] |
| STO-Gr | Terraced substrate | Magnetic sensor | Method b | MR ~ 5000% | [94] |

Note: The fabrication methods are labelled according to the classification presented in Figure 2. The abbreviations are listed as follows. MOSFET, metal-oxide-semiconductor field-effect transistor. EOT, equivalent oxide thickness. $I_{on}/I_{off}$, current on/off ratio. SS, subthreshold swing. CET, capacitance equivalent thickness. Gr, graphene. PLD, pulsed laser deposition. FeFET, ferroelectric field-effect transistor. $I_{ph}$, photocurrent. FTJ, ferroelectric tunnel junction. TER, tunneling electroresistance. UV, ultraviolet. $V_{oc}$, open-circuit voltage. GMR, giant magnetoresistance. MR, magnetoresistance.

**MOSFETs**

2DLMs, in particular semiconducting TMDCs, are considered as a potential channel material for ultra-scaled MOSFETs.[16] The scaling limit of the channel length ($L_g$) is determined by the "natural length"

$$\lambda = \sqrt{t_s \varepsilon_s t_{ox}/\varepsilon_{ox}} \tag{1}$$

where $t_s$, $\varepsilon_s$, $t_{ox}$, and $\varepsilon_{ox}$ are the thicknesses and dielectric constants of the semiconductor channel and gate insulator, respectively.[3, 95, 96] Beyond this limit, which is usually several times of $\lambda$, the transistor tends to suffer from the notorious short-channel effects and lose electrostatic control from the gate. Since $\lambda$ is proportional to the root square of $t_s$, it is highly desirable to reduce the channel thicknesses. In this regard, the innate atomic thickness of single-layer TMDCs is particularly beneficial for the ultimately scaled channel length. In addition, it is viable to retain the high mobility of TMDCs in the single-layer limit thanks to their dangling-bond-free, atomically flat surfaces. In contrast, the mobilities of traditional 3D semiconductors such as silicon would drop substantially when thinned to such thicknesses due to carrier scattering by thickness fluctuation, surface roughness and dangling bonds.[97, 98] Because of these attributes of 2DLMs, the 2022 International Roadmap for Devices and Systems (IRDS[TM]) lists them as a potential channel material for continued scaling-down of transistor sizes.[4]

According to the rule described by Equation (1), the scaling limit of channel lengths is inversely proportional to the dielectric constant of the gate insulator. Thus, a large dielectric constant of the gate insulator is also beneficial for minimizing the channel lengths. In addition, since the operation of MOSFETs is based on the modulation of the channel conductance through the MOS capacitor (Figure 3a), the gate insulator with a large dielectric constant is also beneficial to reducing the required gate voltages, which is directly related to the power consumption. For these reasons, tremendous efforts have been made to integrate high-$\kappa$ dielectrics with traditional silicon channel as well as with 2DLMs.[99-102] STO, a quantum paraelectric perovskite oxide, exhibits an exceptionally large dielectric constant in the range of 200–300 at room temperature, which is appealing for gate dielectric applications.[103, 104] Attracted by such a high dielectric constant, Kim *et al.* fabricated monolayer $MoS_2$ FETs using ~100-nm-thick STO film as the dielectric layer, which was epitaxially grown on conductive Nb-doped STO (Nb:STO) substrates.[73] They achieved a current on/off ratio of over $10^7$ at a gate voltage span of about ±5 V. Meanwhile, following a few studies on the transport properties of graphene on bulk STiO substrates,[41, 105] Park *et al.* explored graphene-STO thin film heterostructures and demonstrated a reduction of needed gate voltages to ~1 V to observe the quantum Hall effect, from ~10 V in the bulk STO counterparts.[106] These early demonstrations proved the feasibility of using high-$\kappa$ perovskite oxide thin films as the dielectric materials for 2DLM-based FETs, but they are limited to the bulk oxide substrates, which is hardly considered a technologically relevant platform.

Recently, several research groups adopted the freestanding STO membrane technology to fabricate 2D transistors on silicon substrates.[74, 75] Yang *et al.* reported a vdW integration process for STO top-gated $MoS_2$ and $WSe_2$ transistors. Benefiting from the large dielectric constant of STO and the superior interface quality, subthreshold swing (SS) as low as 66 mV dec$^{-1}$ was obtained from the $MoS_2$ MOSFETs (Figure 3b). Huang *et al.* systematically investigated the leakage current through the freestanding STO as shown in Figure 3c. With a thickness of 40 unit cells (u.c.), or ~16.4 nm of which capacitance equivalent thickness (CET) is smaller than 1 nm, the leakage current is far below the low-power limit of $1.5 \times 10^{-2}$ A cm$^{-2}$. This low leakage contradicts the very small band offsets between STO and $MoS_2$ but can be attributed to the van der Waals gap which alleviates tunneling problems. The scaling potential of the $MoS_2$ MOSFETs employing STO as the dielectric layer was evaluated by measuring short-channel transistors. As shown in Figure 3d, the SS values remain low for sub-100-nm channel length. In particular, the 35-nm short-channel $MoS_2$ transistor still exhibits a SS value of 79 mV dec$^{-1}$. These results prove that freestanding STO is a promising dielectric to be integrated with semiconducting TMDCs for ultrascaled MOSFETs. The potential of STO dielectric in CMOS technology based on TMDCs was assessed by measuring a CMOS inverter, which was constructed by connecting a p-type $WSe_2$ transistor in series with an n-type $MoS_2$ transistor (Figure 3e). The output logic, represented by the output voltage ($V_{OUT}$), is opposite to the input logic, represented by the input voltage ($V_{IN}$). The peak voltage gain in the inversion action is considerably larger than unity. Thus, this prototype inverter is suitable for scaling up to achieve more complex logic functions.

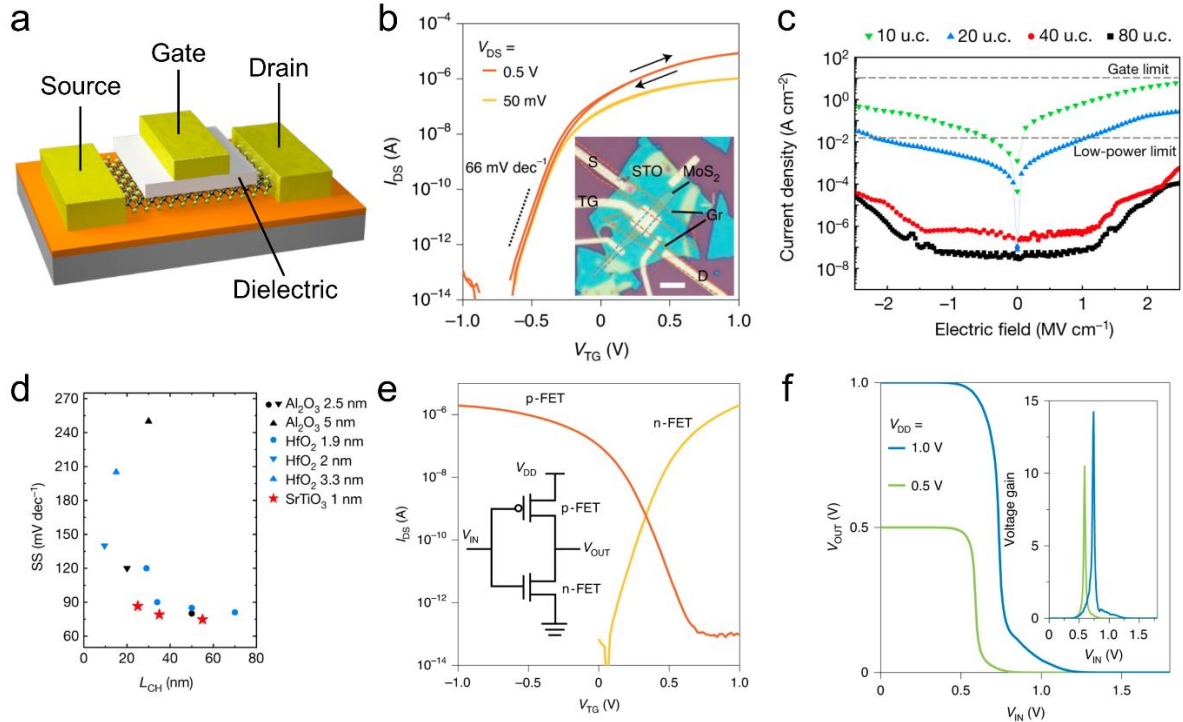

**Figure 3. 2DLM-based high-performance MOSFETs with high-*κ* perovskite oxides as the dielectric layer.** (a) Schematic of a typical MOSFET structure based on semiconducting TMDCs. (b) Transfer characteristics ($I_{DS}$-$V_{TG}$) of a MoS$_2$ MOSFET with STO dielectric layer. Inset, the optical image of the device. Scale bar, 10 μm. (c) Leakage current through STO with different thicknesses measured in unit cells (u.c.). (d) Benchmarking of SS of short-channel (≤100 nm) 2DLM-based MOSFETs with capacitance equivalent thickness (CET) smaller than 5 nm. (e) $I_{DS}$-$V_{TG}$ curves of the p-type WSe$_2$ and n-type MoS$_2$ MOSFETs with STO dielectric layers. Inset, circuit diagram of a CMOS inverter constructed from the p- and n-type MOSFETs. (f) Voltage transfer curves the inverter shown in (e). Inset, voltage gain of the inverter. Panels (b), (e), and (f) are reproduced with permission under a Creative Commons Attribution 4.0 International License from ref 74. Copyright 2022 Springer Nature. Panels (c) and (d) are reproduced with permission from ref 75. Copyright 2022 Springer Nature.

Besides STO, there are a few other high-*κ* perovskite oxides that may be used as gate dielectrics in 2DLM-based MOSFETs. The room-temperature dielectric constant and band gap of these candidate dielectrics, along with those of the commonly used dielectric materials (SiO$_2$, Al$_2$O$_3$, and HfO$_2$), are summarized in Table 2. Some of them, *e.g.* LaAlO$_3$, were considered a promising dielectric material to replace SiO$_2$ during the transition to high-κ dielectrics of Si MOSFETs.[104, 107] Although HfO$_2$ won this competition, partially due to its stability under the high processing temperature, it is meaningful to explore their application in 2DLM-based MOSFETs because their processing conditions are dramatically different. For instance, high-temperature annealing to activate implanted dopants after MOSFET fabrication, which could lead to interlayer diffusion or oxidation of the semiconductor channel, is not required in 2DLM-based MOSFETs. Additionally, the aforementioned freestanding oxide technology can be applied to these high-κ perovskite oxides, thereby adding more flexibility to the fabrication process.

**Table 2.** Room-temperature dielectric constant and band gap of the representative high-κ perovskite oxides.

| Material | Dielectric constant | Band gap (eV) | References |
|---|---|---|---|
| SiO$_2$ | 3.9 | 9.0 | |
| Al$_2$O$_3$ | 9 | 7.0 | [108] |
| HfO$_2$ | 25 | 5.7 | [108, 109] |
| LaAlO$_3$ | 20−27 | 5.6 | [110, 111] |
| BaHfO$_3$ | ~38 | 6.1 | [112] |
| SrHfO$_3$ | ~19 | 6.1−6.5 | [113, 114] |
| LaInO$_3$ | 39 | 5.0 | [115] |
| KTaO$_3$ | 239 | 3.6 | [116, 117] |
| CaTiO$_3$ | 168 | 3.8−4.4 | [118, 119] |
| SrTiO$_3$ | 200−300 | 3.3 | [120, 121] |
| BaZrO$_3$ | 41 | 5.3 | [122, 123] |
| SrZrO$_3$ | 25 | 5.6 | [124] |

The performance of the MOSFETs is significantly affected by the dielectric-semiconductor interfacial properties. The interfacial states trap the free carriers induced by the gate voltage and degrade the SS. This effect can be expressed by the following equation

$$SS = (\ln 10)\left(\frac{k_B T}{q}\right)\left(1 + \frac{C_t}{C_i}\right) \tag{2}$$

where $k_B$, $T$, $q$, $C_t$ and $C_i$ are Boltzmann constant, absolute temperature, elementary charge, areal capacitance related to the trap density, and the areal capacitance of the gate insulator, respectively.[125] With more interfacial states, the SS deviates more from the thermionic limit, 60 mV dec$^{-1}$. These interfacial states also lead to the instability MOSFETs, manifested by the hysteresis of the transfer characteristics.[67, 126] Furthermore, the charged traps scatter the channel free carriers and lead to mobility degradation.[127] Thus, elaborate interface engineering, such as dangling bond passivation and buffer layer growth, should be carried out to optimize the interfacial quality.[128, 129] However, such interfacial imperfections, as well as the bulk defects, with slow or even nonvolatile dynamics, could be exploited for other electronic devices including memory and neuromorphic devices.

**Nonvolatile Memory Devices**

Nonvolatile, high-speed, high-density, and low-power memory devices have long been pursued because they can substantially improve the performance of processing and storing information.[130, 131] Thanks to their atomic thicknesses and diverse electronic properties, 2DLMs offer many possibilities for memory devices working on various mechanisms, including filament forming, charge-trapping, and ferroelectric switching.[132-136] Dependent on the working mechanisms, a variety of corresponding device architectures have been adopted. Among them, ferroelectric field-effect transistors (FeFETs) are a promising platform to exploit the synergistic coupling between 2DLMs and ferroelectric perovskite oxides for memory devices.[28-30, 137] Compared with other ferroelectric materials, perovskite-oxide-based ferroelectric materials, such as PZT and PMN-PT, are praised for their large polarization strength, fast switching speed, and environmental stability.[25] Specifically, the Curie temperature of the perovskite-type ferroelectrics are mostly considerably higher than room temperature, which allows a wide temperature window for the device operation. The remnant polarization are much higher than the organic counterpart, *e.g.* polyvinylidene fluoride (PVDF), and exert a stronger electric field to the adjacent materials. This is due to the extremely large dielectric constant of perovskite-type ferroelectrics. In addition, the high-quality ferroelectric perovskite oxides have a steep switching in the polarization-field hysteresis loop. This is beneficial for nonvolatile memory devices because the binary states can be defined definitely. However, it is worth noting that the ferroelectric properties of perovskite oxides, *i.e.* Curie temperature, remnant polarization, and coercive field, are highly dependent on the growth conditions, composition of the solid solutions, and thickness of the thin films.[25, 138]

FeFETs employ a similar structure to MOSFETs, but the gate dielectric layer is replaced by a ferroelectric material (Figure 4a). The polarization of the ferroelectric perovskite oxides is electrically switchable and retained in the absence of the applied electric field (Figure 4b). The remnant polarization exerts an electric field effect to the adjacent 2DLMs and thus induces or depletes the charge carriers in 2DLMs. This change can be read out by measuring the current flowing through the 2DLM channel. Despite the apparently simple working mechanism, early studies did not observe the anticipated polarization-induced hysteresis.[79, 139] Rather, charge-trapping dominated the interface and led to a hysteresis of which direction was opposite to the ferroelectric polarization. In view of this, Lipatov *et al*. proposed a special electrical measurement scheme that eliminated the interference of charging trapping.[80] Using this measurement scheme, they successfully revealed the ferroelectric polarization-induced $I_{DS}$-$V_G$ hysteresis as shown in Figure 4c. In this way, they measured the improved ON/OFF window and retention properties of the graphene FeFET (Figure 4d). For practical application, however, it is desirable to directly control the channel conductance using ferroelectric polarization. To this end, Lu *et al*. grew single-crystalline PZT on lattice-matched, SRO-buffered STO substrates to fabricate $MoS_2$ FeFETs.[140] They compared the PZT films with smooth and rough surfaces and concluded that the smooth surface of PZT is crucial to achieve genuine FeFETs.

The polarization switching of ferroelectric perovskite oxides can also be harnessed for memory devices using a tunnel structure called ferroelectric tunnel junctions (FTJs).[141] In this structure, the thin ferroelectric perovskite oxide film is sandwiched between two electrodes. The tunneling barrier, hence the overall resistance, of the junction is intimately related to the polarization of the ferroelectric layer. Li *et al.* used semiconducting $MoS_2$ as one of the electrodes to boost this tunneling electroresistance (TER) effect.[83] As shown in Figure 4e and f, the direction of the BTO polarization dramatically alters the band alignment of the $MoS_2$-BTO-SRO junction, as well as the carrier properties of $MoS_2$. Compared with the graphene,[82] $MoS_2$ exhibited a larger change of conductivity with the polarization reversal of the adjacent ferroelectric material. Thus an ON/OFF ratio as large as $10^4$ was achieved. Similar to the FeFETs, the memory performance of FTJs is also strongly affected by the interfacial properties. As shown in Figure 4h, the adsorbed $NH_3$ molecules tend to stabilize the polarization of the BTO whereas $H_2O$ molecules destabilize it.

On one hand, the charge traps and adsorbed molecules at the interfaces of 2DLM-perovskite oxides obscure the ferroelectric polarization, which is usually detrimental to the operation of FeFETs and FTJs. On the other hand, these trap states can also be exploited for memory devices, especially when coupled with light illumination. First, photo-generated charge carriers captured by these trap states may have a long life time. They exert a long-lasting photogating effect on the 2DLMs, thus dramatically changing their conductance.[46, 86, 142] Second, the accumulated charges at the surface may depolarize the ferroelectric materials, thereby restoring the electronic properties of the adjacent 2DLMs.[51, 143] These processes can be used to optically write or erase the memory states.[78, 85] For example, as shown in Figure 4i, two operation schemes, *i.e.* "optical erase-electrical write" and "electrical erase-optical write", can be implemented on the $MoS_2$-PZT memories.

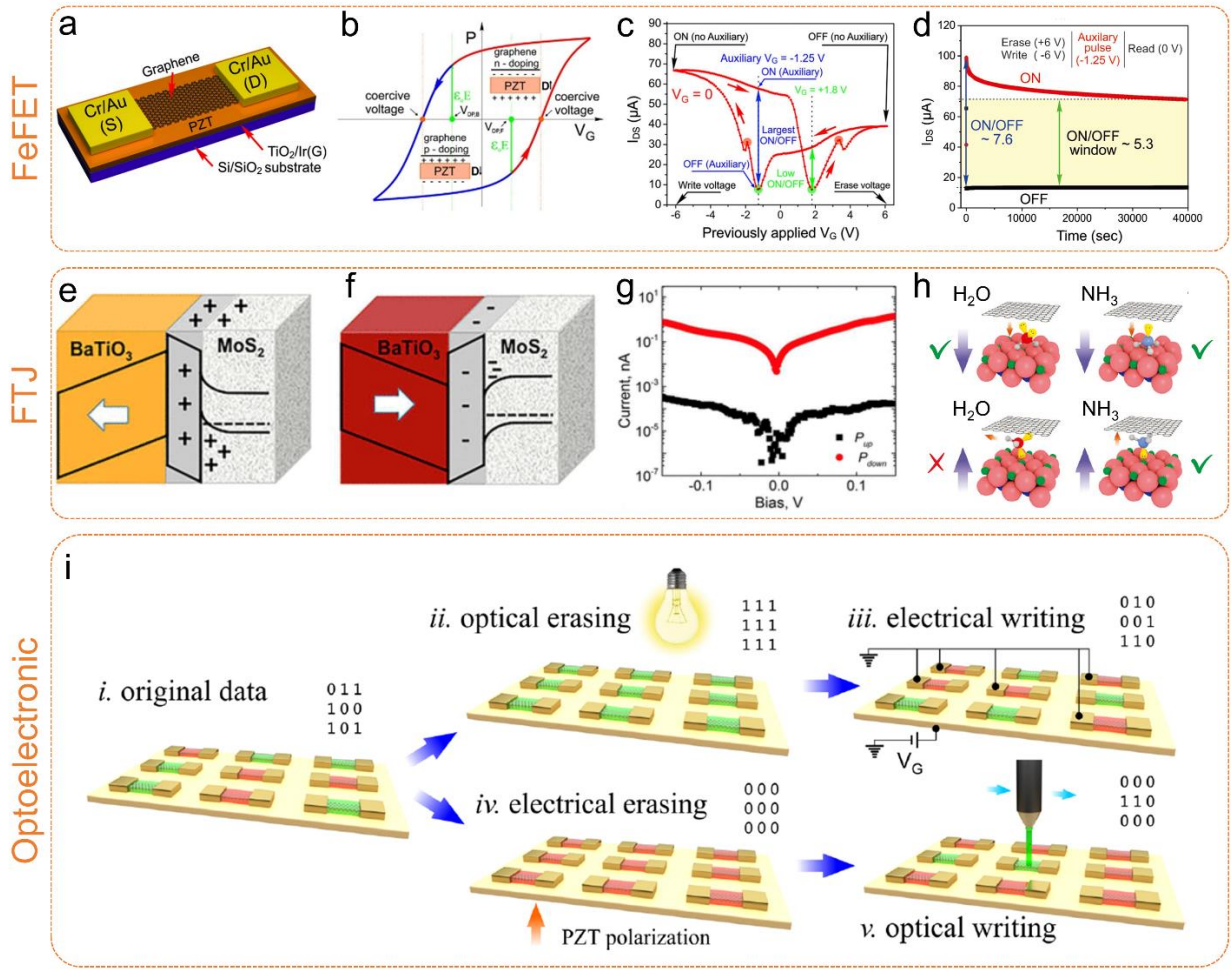

**Figure 4. Nonvolatile memory devices based on 2DLM-provskite oxide heterostructures.** (a) Schematic of a graphene FeFET with PZT as the ferroelectric layer. (b) Polarization ($P$) versus gate voltage ($V_G$) of the graphene FeFET. (c) The ferroelectric-polarization-induced hysteretic $I_{DS}$-$V_G$ curve. (d) Retention characteristics of the FeFET measured by minimizing the charge trapping. Band structures of the MoS$_2$-BTO-SRO FTJ with (e) downward and (f) upward BTO polarization. (g) $I$-$V$ curves of the MoS$_2$-BTO-SRO FTJ with opposite BTO polarization. (h) Stable and unstable polarization of BTO interfaced with NH$_3$ and H$_2$O molecules, respectively. (i) A scheme of optoelectronic memory devices based on MoS$_2$-PZT heterostructures using electrical and optical writing and erasing. Panels (a-d) are reproduced with permission from ref 80. Copyright 2017 John Wiley & Sons. Panels (e-g) are reproduced with permission from ref 83. Copyright 2017 American Chemical Society. Panel (h) is adapted and reproduced with permission

from ref 82. Copyright 2014 Springer Nature. Panel (i) is reproduced with permission from ref 51. Copyright 2015 American Chemical Society.

**Neuromorphic Devices**

Neuromorphic computing using a brain-inspired computing paradigm aims to overcome the limitations faced by the conventional von Neumann architecture.[144-146] It is characterized by processing the information in the memory, thereby obviating the shuttling of data between the processing unit and memory. The hardware implementation of neuromorphic computing exploits both nonvolatile and volatile memory devices for different applications.[147] The gradual decay of conductance in volatile memory devices can be used to mimic the behavior of biological synapses while multi-level nonvolatile memory devices allow implementation of vector-matrix multiplication with high efficiency. Perovskite oxides exhibit a range of effects with different dynamics that can be harnessed to fabricate such devices. Besides the trapped charges at the interface, the oxygen vacancies in the perovskite oxides also exert an electric field effect to the surroundings. These oxygen vacancies migrate under an electric field, thereby altering their electrostatic modulation to the adjacent semiconductor channel (Figure 5a).[70, 84, 87] In a FeFET employing $Bi_4Ti_3O_{12}$ and $MoS_2$ as the ferroelectric and channel material, respectively, Wang *et al.* used this effect to demonstrate an artificial synapse. As shown in Figure 5b, the synaptic plasticity is mimicked by quasi-nonvolatile response of postsynaptic current (PSC), which is represented by the drain-source current, to the presynaptic voltage pulses applied to the gate. More significantly, when two consecutive gate voltage pulses were applied at small time intervals, the PSC was largely enhanced (Figure 5c and d), a phenomenon known as paired-pulse facilitation (PPF). These results clearly proved the feasibility of using 2DLM-perovskite oxide heterostructures for neuromorphic devices.

Du *et al.* explored the 2D semiconductor/ferroelectric optoelectronic transistor for neuromorphic vision sensors.[88] The FeFETs built on the $MoS_2$ channel and BTO ferroelectric layer (Figure 5e) displayed wavelength-dependent nonvolatile photoresponse. As shown in Figure 5f, the conductance of $MoS_2$ is increased by light illumination for all wavelengths and persists after turning off the light. However, the photoresponse under 450 nm illumination is significantly larger than 532 and 650 nm. This selectivity on wavelength, combined with the quasi-nonvolatile photoresponse, was used for pre-processing of visual information (Figure 5g). The color information of the photograph was mapped to the intensity of light with corresponding wavelengths and fed to the FeFET. After a certain number of light pulses, the feature of the object with the most blue component emerged from the background. Based on the simulation, they also found that after such preprocessing, the recognition rate of handwritten numbers can be dramatically improved.

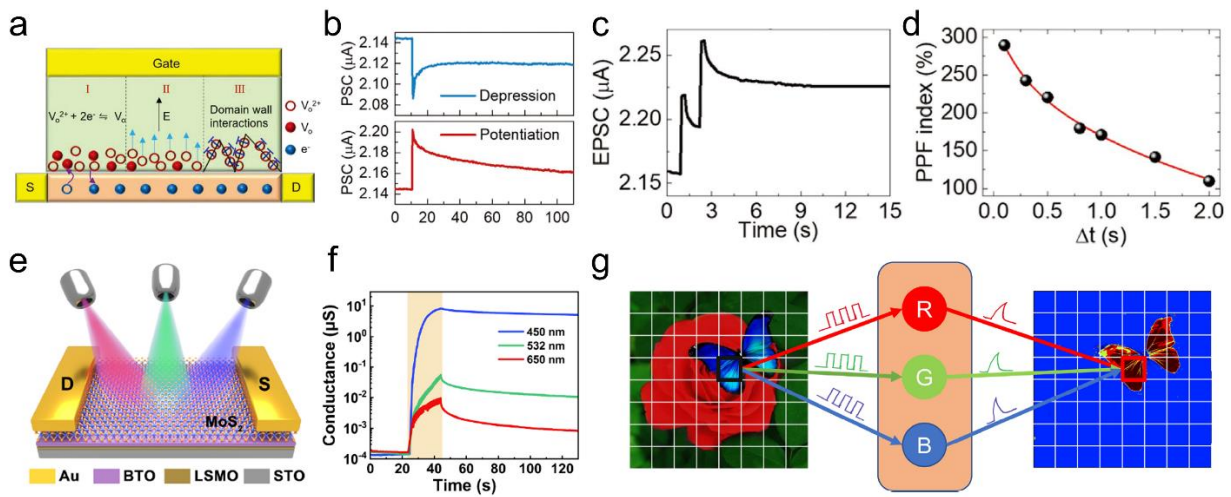

**Figure 5. Neuromorphic devices based on 2DLM-perovskite oxide heterostructures.** (a) Schematic illustration of dynamic processes of charges and oxygen vacancies in a $MoS_2$ FeFET. The charge trapping and releasing are indicated by the purple arrows. The drift of oxygen vacancies under the electric field is denoted by the blue arrows. (b) Short-term depression and potentiation revealed by the post-synaptic current (PSC) in response to pre-synaptic voltage pulses. (c) Paired-pulse facilitation (PPF) demonstrated by the enhanced PSC due to two consecutive pre-synaptic voltage pulses. (d) PPF index, the ratio of the second PSC peak to the first, as a function of the time interval between two consecutive pre-synaptic voltage pulses. (e) Schematic of the neuromorphic vision sensor comprising the $MoS_2$ channel and the BTO ferroelectric layer in a FeFET configuration. (f) The change of $MoS_2$ conductance under 100 pulses of light illumination of different wavelengths. The light intensities are all 10 mW/cm$^{-2}$. (g) Schematic of extracting color information from a photograph by measuring varied photoresponse to different

wavelengths. Panels (a-d) are reproduced with permission from ref 87. Copyright 2022 John Wiley & Sons. Panels (e-g) are reproduced with permission from ref 88. Copyright 2021 Elsevier.

**CONCLUSIONS AND OUTLOOK**

The current status of electronic devices based on 2DLM-perovskite oxide heterostructures is presented in this Review. We have outlined the potentially technologically important properties of perovskite oxides and how they can be integrated with 2DLMs. The interactions between 2DLMs and perovskite oxides lead to prototype electronic devices ranging from high-performance MOSFETs, nonvolatile memory devices, to neuromorphic devices. These advancements represent the first and crucial step in using 2DLM-perovskite oxide heterostructures as the building block for future information processing technology. The strategies used in these heterostructures can be extended to more complex device structures and other material combinations. For example, negative-capacitance (NC) FETs that employ bilayer dielectrics comprising a ferroelectric layer and a dielectric layer may be realized to further reduce the SS. The layer-by-layer stacking method should be advantageous to capacitance matching because the individual ferroelectric and dielectric layers can be optimized before integration. These strategies to integrate perovskite oxides could also be applied to other material families, such as the halide perovskites.[148, 149] Unlike the usually insulating perovskite oxides, the halide perovskites are typical semiconductors with outstanding optoelectronic properties and have a great potential in solar cells, light-emitting diodes, and photodetectors. Integration of high-$\kappa$ perovskite oxides with the halide perovskite may boost the performance of the halide perovskite electronics.

Currently, the 2DLM-perovskite oxide heterostructures for electronic devices exploit mostly dielectric or ferroelectric properties of perovskite oxides. Although magnetic proximity effects between 2DLMs and perovskite oxides have been intensively investigated, the potential of these effects for device applications is rarely studied.[26] With the help of freestanding oxide technology, we envision the fabrication of complex structures from magnetic perovskite oxides and 2DLMs to realize various spintronic devices, including spin valves, spin FETs, and spin-orbital coupling (SOT) devices.[21, 150] Based on the similar rationale, a wide range of exotic properties of perovskite oxides, such as the high-$T$c superconductor and piezoelectricity, can be integrated with 2DLMs for advanced device applications, such as Josephson junctions, nanoelectromechanical devices.[151]

However, to realize practical applications, a few challenges remain to be overcome.

The functionalities of 2DLM-perovskite oxide heterostructures are strongly affected by a range of intrinsic and extrinsic factors, including surface roughness and crystal defects of perovskite oxides, and charge traps and adsorbates at the interfaces. These factors complicate the interactions between 2DLMs and perovskites, and make the fundamental study of physical and electronic properties difficult. They may dominate the physical process, thus obscuring the interactions of interest. From the application point of view, these factors undermine the usefulness of the heterostructures when they have opposite effects on some physical processes. Therefore, considerable attention should be paid to improving the material quality as well as surface modification. For example, post-growth annealing of perovskite oxides has been proven a viable approach to minimizing oxygen vacancies in perovskite oxides. A thin layer of hexagonal boron nitride (hBN), a surface-inert insulator, may be inserted between 2DLMs and perovskite oxides to mitigate the charge-trapping problem.[152, 153]

For potential practical applications, the devices based on 2DLM-perovskite oxide heterostructures should be thoroughly characterized in terms of the dynamic performance and reliability. Specifically, the switching speed of the MOSFETs should be measured and benchmarked with the current Si technology and other emerging material combinations. The variation of the device characteristics over time requires further experimental and theoretical study.[154] In addition, wafer-scale transfer of freestanding perovskite oxides is highly desirable for large-scale integration and device fabrication. Based on these achievements, the electronic devices based on 2DLM-perovskite oxide heterostructures are expected to provide many high-performance and unconventional functionalities that fuel the next-generation information technology.

**Acknowledgements**

X.R.W. acknowledges support from Singapore Ministry of Education under its Academic Research Fund Tier 2 (grant nos. MOE-T2EP50120-0006 and MOE-T2EP50220-0005) and Tier 3 (grant no. MOE2018-T3-1-002); and the Agency for Science, Technology and Research (A*STAR) under its AME IRG grant (project no. A20E5c0094).